\documentclass{article}

\usepackage{PRIMEarxiv}

\usepackage[utf8]{inputenc} 
\usepackage[T1]{fontenc}    
\usepackage{hyperref}       
\usepackage{url}            
\usepackage{booktabs}       
\usepackage{amsfonts}       
\usepackage{nicefrac}       
\usepackage{microtype}      
\usepackage{lipsum}
\usepackage{fancyhdr}       
\usepackage{graphicx}       
\usepackage{arydshln}
\usepackage{multirow}
\graphicspath{{media/}}     

\title{Explainable Techniques for Analyzing Flow Cytometry Cell Transformers
}

\author{
  Florian Kowarsch, Lisa Weijler, Florian Kleber \\
  Computer Vision Lab, TU Vienna \\
  \And
   Matthias Wödlinger, Michael Reiter \\
   Computer Vision Lab, TU Vienna \\
   St. Anna Children's Cancer Research Institute \\
   \And
  Margarita Maurer-Granofszky, Michael Dworzak \\
  St. Anna Children's Cancer Research Institute \\
  Labdia Labordiagnostik GmbH, Vienna \\
  \texttt{dworzak@ccri.at} \\
}

\begin{document}
\maketitle

\begin{abstract}
 Explainability for Deep Learning Models is especially important for clinical applications, where decisions of automated systems have far-reaching consequences.
  While various post-hoc explainable methods, such as attention visualization and saliency maps, already exist for common data modalities, including natural language and images, little work has been done to adapt them to the modality of Flow CytoMetry (FCM) data.
  In this work, we evaluate the usage of a transformer architecture called ReluFormer that ease attention visualization as well as we propose a gradient- and an attention-based visualization technique tailored for FCM.
We qualitatively evaluate the visualization techniques for cell classification and polygon regression on pediatric Acute Lymphoblastic Leukemia (ALL) FCM samples. 
The results outline the model's decision process and demonstrate how to utilize the proposed techniques to inspect the trained model.
The gradient-based visualization not only identifies cells that are most significant for a particular prediction but also indicates the directions in the FCM feature space in which changes have the most impact on the prediction.
The attention visualization provides insights on the transformer's decision process when handling FCM data.  We show that different attention heads specialize by attending to different biologically meaningful sub-populations in the data, even though the model retrieved solely supervised binary classification signals during training.
\end{abstract}


\section{Introduction}

\begin{figure}
    \centering
    \includegraphics[width=0.99\linewidth]{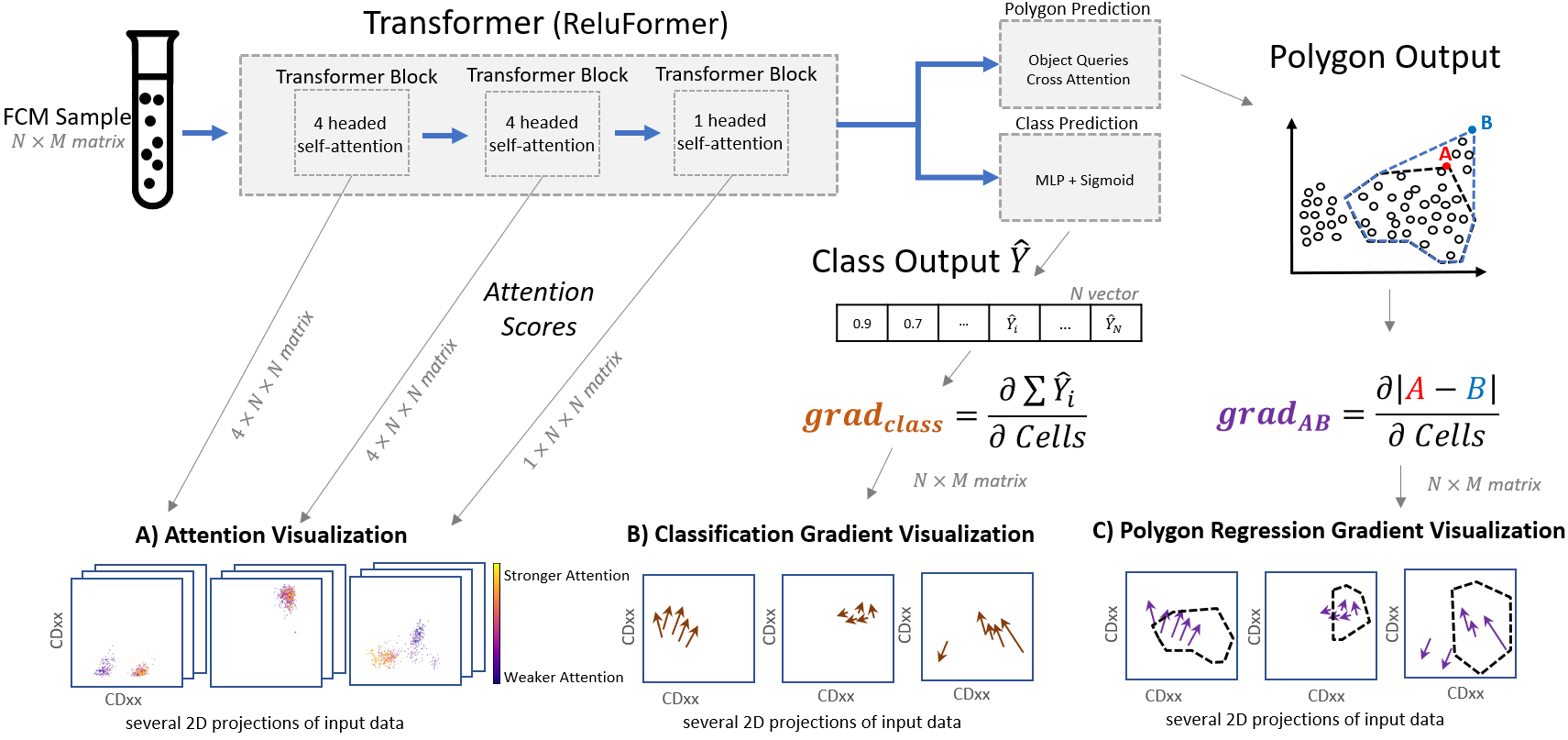}
    \caption{A) Attention scores are retrieved from the intermediate layers of the model. We plot the top 500 events with strongest attention scores in different 2D projections of the data. B) The gradients of the summed event-wise class predictions with respect to the input data are computed. The  top 100 gradient vectors are displayed, which  point in the direction of fastest change of the predicted class. C) The gradients of the distance between the predicted polygon to a constructed query polygon. We depict the vectors of the top 100 gradients, which indicate the direction the input data should change to minimize the difference between predicted and query polygon.}
    \label{fig:visual_abstract}
\end{figure}

Deep learning models have wide applications to problems in a clinical context, see \cite{piccialli2021survey} for a comprehensive survey. However, since predictions of these models can have far-reaching consequences, it is important that these predictions are transparent and interpretable. This is also true in the field of automated cell detection in flow cytometry (FCM) data. FCM is used to measure the expression levels of antigens on blood or bone marrow cells and is commonly used in research and clinical practice for tasks such as immunophenotyping or monitoring residual cancer cells (Measurable Residual Disease, MRD) during chemotherapy. A typical sample contains 50-500k cells, each with up to 15 different features that correspond to the physical properties of the cells or the expression levels of specific antigens. 
To analyze FCM samples in clinical practice, medical experts view two-dimensional projections of the data and distinguish sub-populations of events by drawing polygons called \textbf{gates} around them \cite{mckinnon2018flow}. This hierarchical process is known as \textbf{gating}.
Automated methods either directly predict a class label for each cell (\textbf{classification}) or predict the polygons of the gating procedure (\textbf{polygon regression}).
Current state-of-the-art methods rely on a transformer-based architecture as it allows to process a whole FCM sample at once and thereby accounts for inter-cell relations \cite{wodlinger2022automated}.
While these approaches perform comparable to human experts, they lack ways to analyze their inner structure and interpret their predictions, which, as past has shown \cite{zech2018variable,amann2020explainability}, is essential to ensure that the desired concepts are learned and biased decision are avoided.

Although various explainability \footnote{Following \cite{molnar2020interpretable} we use the terms explainability and interpretability interchangeably.} methods for transformers exist for other data modalities such as natural languages \cite{vig2019multiscale,atanasova2020diagnostic} or images \cite{carion2020end,liu2021visual} they have not been adopted for FCM data. 
Existing explainable AI methods can be divided into \textbf{Intrinsically interpretable}, models that are interpretable due to their internal structures (e.g. decision trees) or \textbf{Post-hoc interpretation}, methods that analyze a model after training \cite{molnar2020interpretable}. Since we aim to enhance the explainability of existing transformer-based FCM models we focus on post-hoc interpretation methods.

A common post-hoc method to interpret a transformer's decisions is to visualize how the model attends to different parts of the input data \cite{bahdanau2014neural,bau2018identifying,vig2019multiscale}, often called \textbf{attention visualization}.
For instance, Jesse Vig \cite{vig2019multiscale} proposed to investigate self-attention between word-tokens by visualizing weighted edges between words and coloring the words based on the attention magnitude.
For computer vision tasks, overlaying the input image with a heatmap is commonly used to visualize the attention \cite{dosovitskiy2020image,aflalo2022vl,carion2020end}.
Attention visualization facilitates model interpretability since it reveals what part of the data and which relationship among the data is considered important by the model. It can be used to verify that the model learned desired concepts or to spot a bias like attending to the background of an image to classify an object in the foreground. 

While attention visualization is a model-specific technique for attention-based architectures, other common post-hoc explainability techniques for deep learning models, which are not restricted to models using attention, include \textbf{gradient based methods} such as Saliency Maps \cite{simonyan2013deep} or Gradient-weighted Class Activation Maps (Grad-CAM) \cite{selvaraju2017grad}.
These and related methods rely on computing the model's gradients of a specific class output with respect to the input image. Pixels with a low gradient norm are considered unimportant for the model's class prediction, and pixels with a high gradient norm are important for the prediction, as changing their values leads to a significant change in the class output. Saliency Maps were initially designed for Convolutional Neural Networks (CNNs) \cite{simonyan2013deep} but are also used for vision-based \cite{liu2021visual,amir2021deep}, mixed modality (vision \& language) \cite{aflalo2022vl} as well as pure language  \cite{atanasova2020diagnostic} transformers.

\paragraph{Contribution}
Our contribution is twofold: 
\begin{itemize}
    \item We propose a \textbf{gradient-based visualization technique} for deep learning models operating on FCM data for cell classification and cell population polygon regression. We show how gradient-based visualization techniques can be used to investigate possible overfitting behaviors in FCM polygon regression.
    \item We suggest to use \textbf{ReluFormer}, an adaption of the CosFormer \cite{qin2022cosformer}, to obtain and \textbf{visualize self-attention scores} in FCM data.
We demonstrate with our attention visualization technique, that different heads specialized to attend to different biologically meaningful cell populations even though they have been training to discriminate cancer cells from normal cells.
\end{itemize}

\section{Methodology}



Since not all recorded observations in FCM are actual individual cells but rather clumped cells, air bubbles, or other not relevant particles, an individual observation is called \textbf{event} \cite{doi:https://doi.org/10.1002/9781118487969.ch2}.
An FCM sample describes a set of \textbf{events} $E \in \mathbb{R}^{N \times m}$, where $N$ defines the number of events ($50-500 \times 10^5$) and $m$ denotes the number of features (typically $10-20$ equivalent to the obtained properties of the cells). Furthermore, $Y \in \{0,1\}^N$ represents the set of true class labels.
For \textbf{cell classification} a model, as proposed in \cite{wodlinger2022automated}, is a function $C: E \to \hat{Y}$ with $C(e|W) \mapsto \hat{y}$, where $W$ represents the learned model weights.
The training process adjusts these model weights to minimize a loss $\mathcal{L}$, which is computed between the prediction $\hat{Y}$ and the ground truth $Y$.
For \textbf{polygon regression} a model, as proposed in \cite{kowarsch2022towards}, is a function $R: E \to P$, where $\hat{P}$ is a predicted set of polygons that represents the \textbf{gates} of the \textbf{gating hierarchy}, used to sub-select cells and detect cancer cells in FCM samples.

\subsection{Gradient-based Visualization}

    When training deep neural networks, gradients of the loss $\mathcal{L}$ are computed with respect to the model weights:
    \begin{equation}
     \textnormal{grad}_W = \frac{\partial \mathcal{L}}{\partial W}
    \end{equation}
    The gradients point in the direction of the steepest ascent, such that a gradient descent based optimizer takes a step in the opposing direction aiming to reduce the loss \cite{goodfellow_chapter4,goodfellow_chapter6}.
    In contrast, gradient-based explainability methods usually compute the gradients of a particular output value with respect to the input data \cite{simonyan2013deep} or intermediate representation \cite{selvaraju2017grad}.
    For instance, the gradients of $\hat{Y}_i$, the binary classification output of the $i^{th}$ event of the network $C$, with respect to the input events $E$ is defined by
    \begin{equation}
     \textnormal{grad}_{E_i} = \frac{\partial C(E)_i}{\partial E} = \frac{\partial \hat{Y}_i}{\partial E}\textnormal{.}
    \end{equation}
    Although the gradients are only computed of the $i^{th}$ classification output with respect to the input sample, we obtain a gradient vector for each event because the class prediction for one event depends on its position and the positions of all other events in the sample.
    Gradient-based explainability methods for images take the norm of gradients of each input pixel and visualize them as a heatmap on top of the original image.
    A high gradient norm indicates that these pixels have a strong impact on the prediction of the corresponding class.
    Instead of the gradient norms for FCM data we use the gradient vectors themselves plotted as vector fields on the 2D plots.
    2D projections are a common way to inspect the high-dimensional FCM data. In clinical practice, this emerged as a common standard to document and analyze FCM data as experienced clinicians can spot different biological phenomena in these plots.
The gradient vectors not only indicate which input events are important for a particular prediction but also reveals the direction in input data space that leads to the greatest change in the prediction.

\begin{figure}
    \centering
    \includegraphics[width=\linewidth]{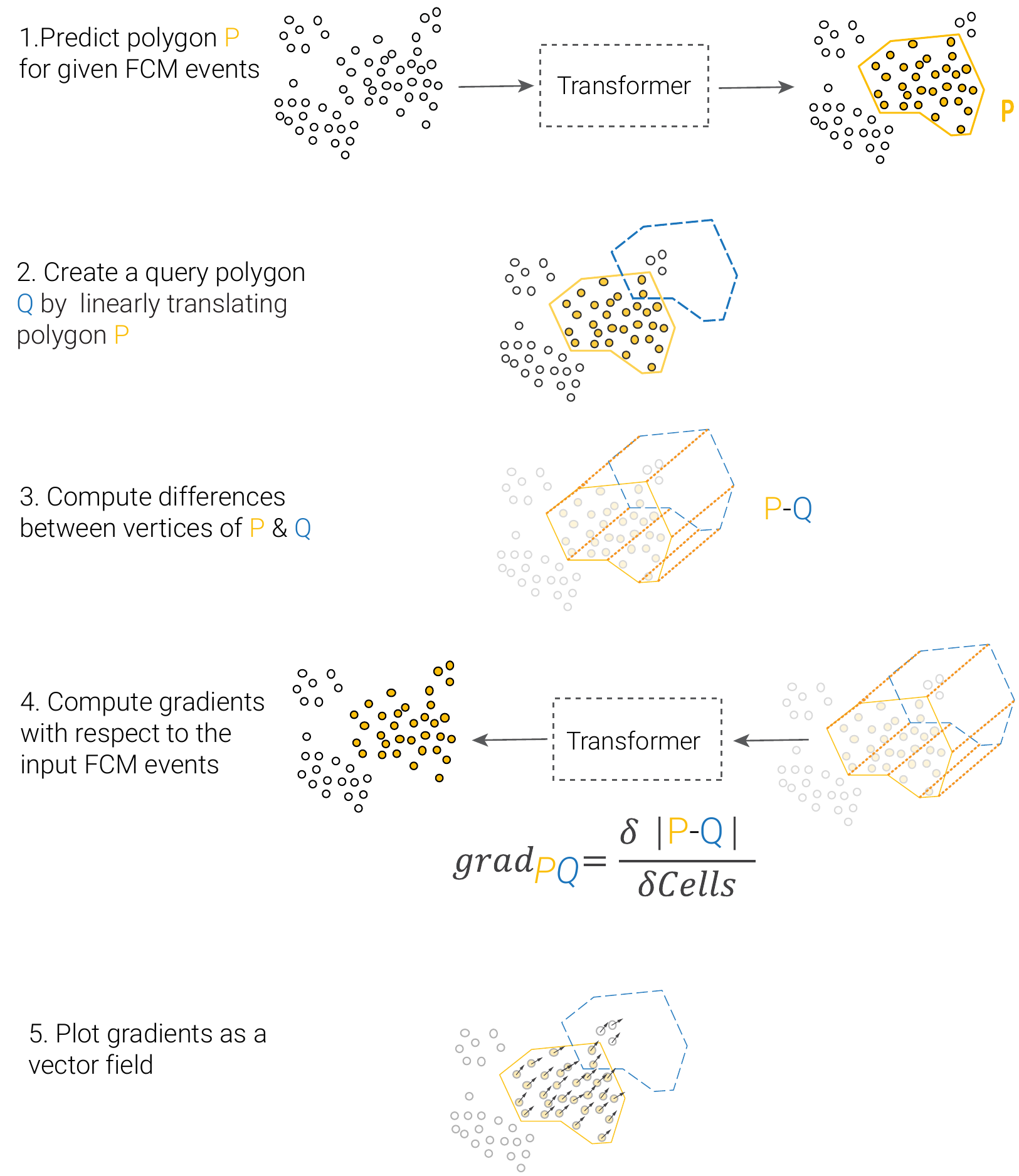}
    \caption{The five steps to visualize the polygon regression gradients for a query polygon Q, addressing the question "How should the input events change in order to predict $Q$ instead $P$?".
    Vectors that do not align with the direction of translation, or vectors that are not positioned inside the predicted polygon, indicate that the model has learned an incorrect relationship between the input data and the predicted polygon.}
    \label{fig:poly_reg_method}
\end{figure}

\paragraph{Polygon Regression Gradients} While above, the gradient-based explainability visualizations are described for classification tasks, we can apply the same concept to polygon regression.
If we define $A$ as the $i^{th}$ vertex of a predicted polygon $P$ 
and $B$ as the actual desired location of that point, then we can calculate the gradients of the norm of $A - B$ with respect to the input events $E$ 
\begin{equation}
     grad_{AB} = \frac{\partial \|R(E)_i - B\|}{\partial E} = \frac{\partial \|A-B\|}{\partial E},
\end{equation} which describes the direction in which the events should shift in order to move the predicted point $A$ to the location of $B$ (as illustrated in Figure \ref{fig:visual_abstract}).
This concept can be extended to not only compute the gradient of differences of one point pair but rather the pairwise difference between vertices of two polygons. Intuitively, this method can be understood as querying a model with a query polygon $Q$ to ask the question "How should the input events change in order to predict $Q$ instead $P$?" Polygon $Q$, for instance, could be a linear translation of $P$. In this case, we expect the gradients to be mainly distributed among the events inside the polygon $P$ pointing in the direction of the translation. Figure \ref{fig:poly_reg_method} illustrates how the gradients are computed and visualized in case of a linearly translated query polygon Q.
    

\subsection{Attention Visualization}

For self-attention the matrix of input tokens $X$ is projected via the learned weight matrices $W_Q, W_K, W_V$ to obtain three matrices Queries (Q), Keys (K) and Values (V).
Then the attention operation is defined as
\begin{equation}
 \textnormal{Attn}(Q, K, V) = \textnormal{softmax}\bigg(\frac{QK^T}{\sqrt{d_k}}\bigg)V, 
\end{equation}
where the \textbf{attention} between Q and K is computed via the dot product, scaled by the square root of the dimension of K ($\sqrt{(d_k)}$) and normalized by the softmax function.
Usually, the result of the softmax-normalized dot product $QK^T$ is used to visualize the attention.
For instance, to visualize the attention score between the $i^{th}$ input token and all other tokens, the result of $Q_iK^T$ can be plotted as a heatmap or weighted graph.
In a transformer model, the attention mechanism is usually computed multiple times per layer, performed by different heads. Common attention visualization tools either allow to switch between different heads of the model \cite{vig2019multiscale} or visualize the last layer's singled-headed attention \cite{aflalo2022vl,carion2020end}.

\paragraph{Attention in FCM Models} 
The memory complexity of the original transformer \cite{vaswani2017attention} is quadratic in the input length $\mathcal{O}(N^2)$ as each input token attends to each other. However, FCM samples can have up to $500 \times 10^5$ events, making them infeasible to process with the original transformer.
To overcome the quadratic complexity Woedlinger et al. \cite{wodlinger2022automated}  used an efficient transformer called \textbf{set transformer} \cite{lee2019set}, which instead of attending between each input token, learns,  similar to vector-quantization \cite{gray1984vector}, a set of $k$ prototype vectors to which the input tokens attend to and thereby reduce the complexity to $\mathcal{O}(N \cdot k)$.
However, this indirect computation of self-attention is undesired for visualization purposes.  
Thus, in this work, we utilize another efficient transformer called \textbf{cosFormer} \cite{qin2022cosformer}
, which overcomes the quadratic complexity by linearization of self-attention.
While the original transformer uses the non-decomposable similarity function $S(Q,K) = \exp(QK^T)$ the authors argue when using a decomposable similarity function such that $S(Q_i,K_j) = \phi(Q_i) \phi(K_j)^T$ we can exploit a matrix product property and compute $\phi(K)^TV$ before we multiple the result with $\phi(Q)$
\begin{equation}
    (\phi(Q)\phi(K)^T)V = \phi(Q)(\phi(K)^TV),
\end{equation}
which avoids the necessity of materializing the $N^2$-sized attention matrix $A=QK^T$.
Qin et al. aims to approximate the main properties of softmax by applying ReLU \cite{nair2010rectified} to ensure non-negativity and a cosine-based re-weighting mechanism that enforces locality.
However, the re-weighting mechanism assumes that the input forms a sequence, which is not given for the set of events forming an FCM sample, we, therefore, omit this part of the cosFormer. In the following, we refer to this simplified version of the cosFormer as \textbf{ReluFormer}.
We observed that compared to the set-transformer ReluFormer performs similar for cell classification on the same experiments as in \cite{wodlinger2022automated}, while allowing to compute attention scores between events for visualization purposes directly.
To obtain attention scores between any events of interest we can simply compute the attention matrix $A_{viz}=ReLU(\check{Q})ReLU(\check{K})^T$, where $\check{Q}$ and $\check{K}$ represents the queries and keys for the selected events respectively and $\check{A}_{viz}$ represents the row-normalized matrix.
 Focusing on the attention of a single event can provide too much detail and thereby miss to depict more sample-wide data relations.
    Often we are interested in inspecting the attention of a particular biological reasonable sub-population. To do so, we can, for instance, sum the attention of all cancer cells to all other events in an FCM sample and visualize the attention scores color-coded on several 2D projections of the high-dimensional data.

\section{Results}
\label{sec:res}

\paragraph{Gradient-based Visualization}

\begin{figure}
    \centering
    \includegraphics[width=0.55\linewidth]{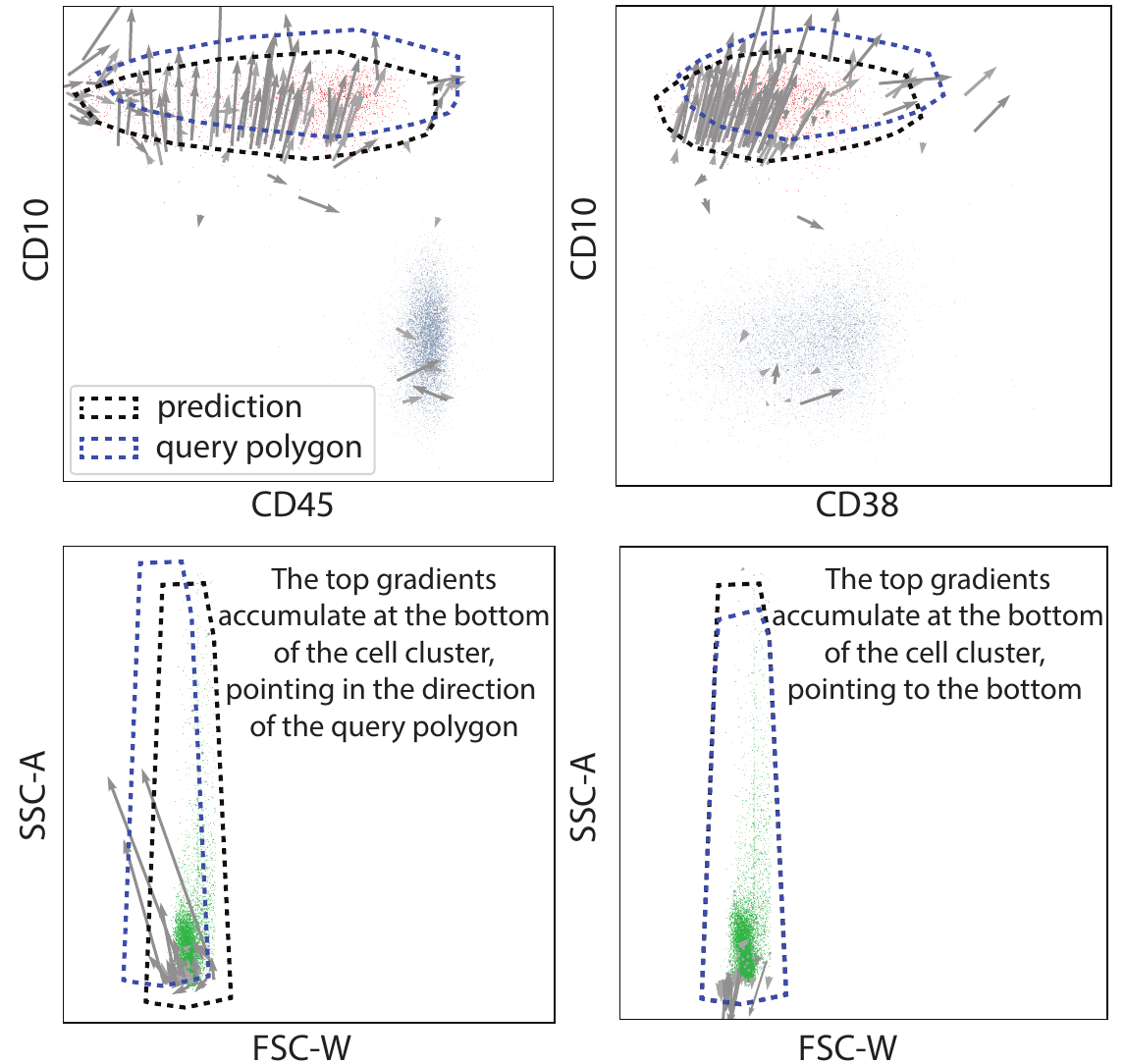}
    \caption{Gradients for Polygon regression can be used to confirm correctly learned relationships (first row) and to spot overfitting behavior (second row). We plot the top 100 gradients of the difference between predicted polygon and query polygon.}
    \label{fig:polygon_shifts}
\end{figure}

Figure \ref {fig:polygon_shifts} demonstrates the proposed gradient-based visualization technique for analyzing an FCM polygon regression model. We first compute the gradients of the difference between the predicted polygon and a slightly transformed \textbf{query polygon}.
Then, we plot the top 100 biggest gradients, which indicate the direction in which input events should change to minimize the difference between the two polygons.
The examples depicted in the first row demonstrate that the model has accurately learned how the position of specific events relates to the gate polygon.
This is because the gradients show that a shift of the predicted polygon is mainly caused caused by a shift of the events inside the polygon in the same direction. However, the prediction for the FCM sample on the bottom row shows that the model has learned an incorrect relationship between the event positions and the polygon position, despite the predicted polygon position being correct. When the polygon is shifted in any direction, the gradients are mainly at the bottom of the event cluster. This suggests that the gradients are not distributed evenly among all events within the polygon and consequentially that the model's prediction of the polygon relies on the position of the cluster bottom.

\paragraph{Attention Visualization}

\begin{figure}[h]
    \centering
    \includegraphics[width=0.83\linewidth, keepaspectratio]{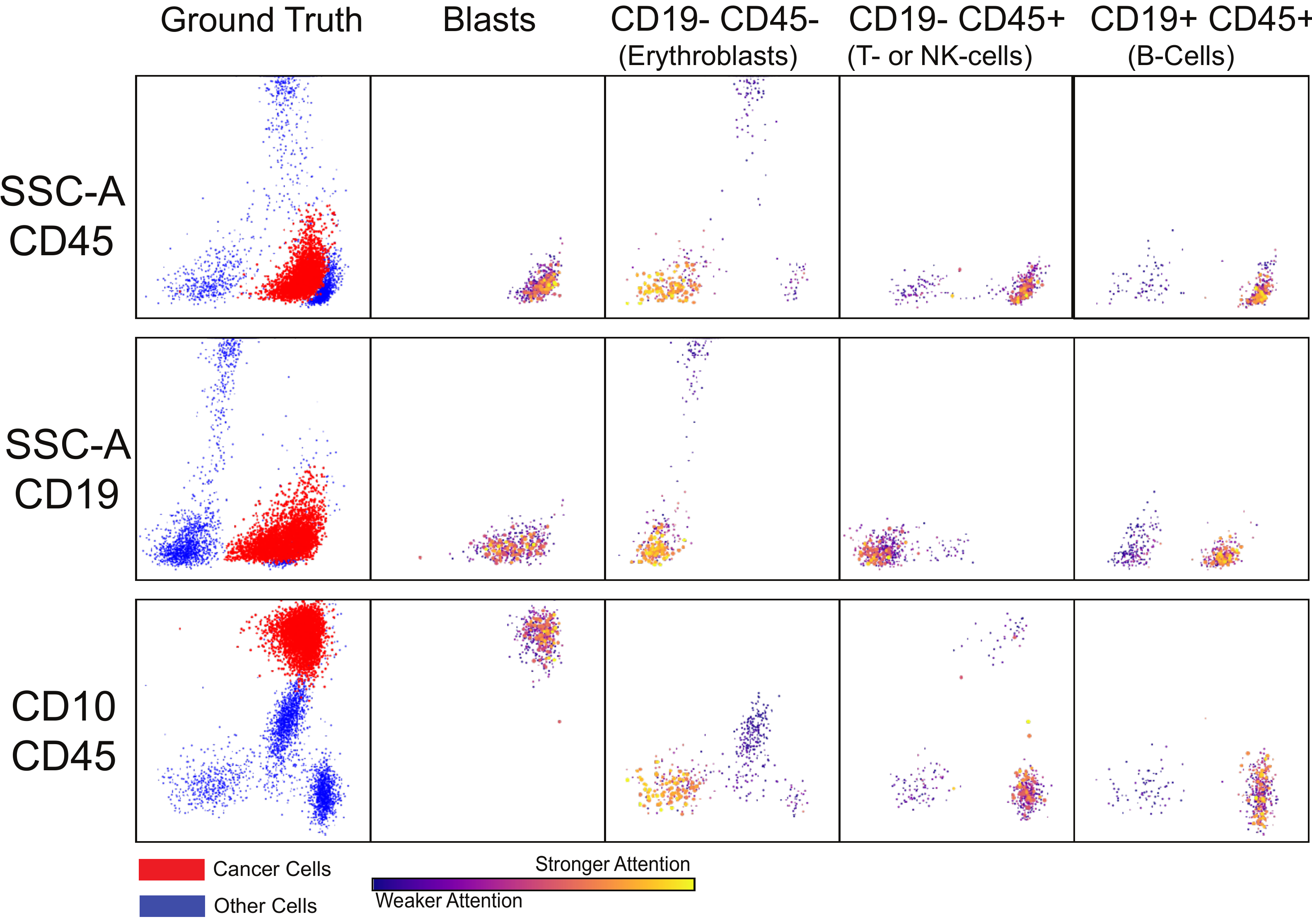}
    \caption{
    Different heads attend to different biological meaningful cell populations.
    Each row shows the same FCM sample from a different 2D projection (e.g. SSC-A/CD45).
    The first column depicts the cells' class-membership. The other columns show the top 500 events with strongest attention  for different heads.}
    \label{fig:atten_heads}
\end{figure}

First, we show in table \ref{tab:comparing_results} that ReluFormer performs comparable to Set-Transformer in experiments. We conduct the same experiments as in \cite{wodlinger2022automated} in binary classification for pediatric B-ALL on 4 different datasets. For a detailed dataset description, the reader is referred to \cite{reiter2019automated} for VIE14, BLN and BUE, and to \cite{wodlinger2022automated} for VIE20.
Figure \ref{fig:atten_heads} visualizes to which events 1000 randomly sampled cancer cells (blasts) of an arbitrarily selected B-ALL sample attend the most. Although the network is trained solely for binary classification (cancer cells vs. non-cancer cells), we can see, that the heads focus on different biological meaningful populations such as CD19- CD45- (likely erythroblasts) or CD19+ CD45+ (likely healthy B-cells).

\begin{figure}[h]
    \centering
\includegraphics[width=0.9\linewidth, keepaspectratio]{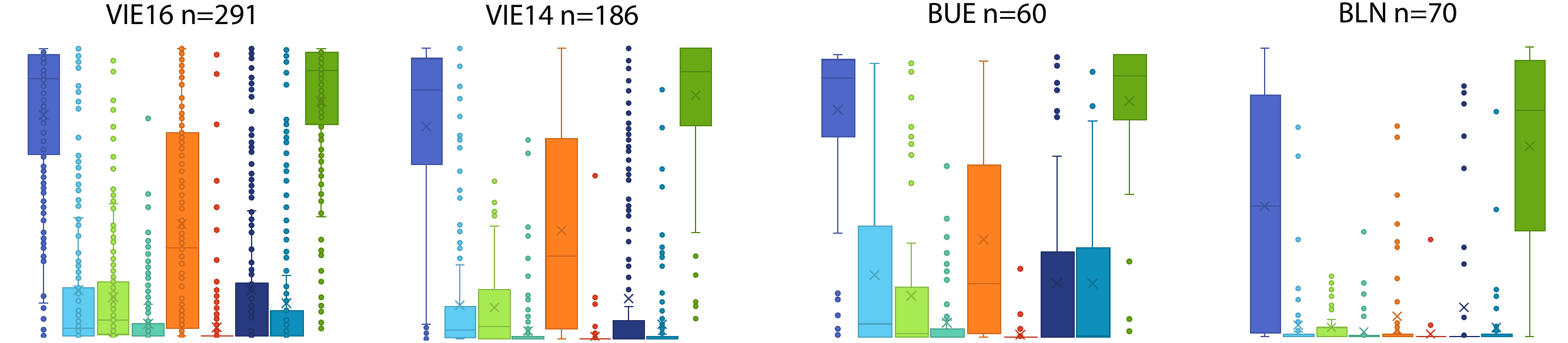}
    \caption{Specific heads (blue and green) show higher amount of cancer cells among the top 5\% attending cells then other heads. }
    \label{fig:quant_res}
\end{figure}

To quantitatively support this observation we first calculate for each FCM sample and each head the top 5\% of cells according the attention (see Figure \ref{fig:quant_res}). Then the amount of cancer cells among these cells are computed the following:
    \begin{equation}
    \label{eg:quant_blast_head}
    \frac{\textnormal{n\_cancer@top5\%}}{min(\textnormal{n\_cells@top5\%},\textnormal{n\_cancer})}    
    \end{equation}
We trained the model for binary cancer classification on VIE14 dataset and calculated the metric on 4 different datasets. Over all datasets the same heads consistently expressed higher amount of cancer cells among the top 5\% attending cells then other heads.
This supports our hypothesis, that specific heads learn to focus on biological meaningful structures. 

\begin{table}
    \centering
    \caption{
    The proposed ReluFormer (adaption of the cosFormer \cite{qin2022cosformer}) compared to
     set-transformer \cite{wodlinger2022automated} on the same experiments as in \cite{wodlinger2022automated} for cancer cell classification. The table reports mean F1-Score / median F1-Score.}
    \begin{tabular}{cc|cccc}
        \textbf{Train}                          & \textbf{Test} & \textbf{Set-Transformer} & \textbf{ReluFormer} \\ \hline \hline
        \multirow{3}{*}{\textbf{VIE14}}         & \textbf{BLN}  & 0.75/0.90 & 0.80/\textbf{0.93} \\
                                                & \textbf{BUE}  & 0.78/\textbf{0.95}  & 0.81/\textbf{0.95} \\
                                                & \textbf{VIE20} & 0.73/0.89 & 0.78/\textbf{0.93}\\ \hline
        \multirow{3}{*}{\textbf{VIE20}}         & \textbf{BLN}  & 0.66/\textbf{0.81} & 0.53/0.58 \\
                                                & \textbf{BUE}  & 0.71/0.86 & 0.72/\textbf{0.91} \\
                                                & \textbf{VIE14} & 0.71/0.86 & 0.83/\textbf{0.94} \\ \hline
        \multirow{3}{*}{\textbf{BLN}}           & \textbf{BUE} & 0.66/\textbf{0.87} & 0.69/\textbf{0.87} \\
                                                & \textbf{VIE14} & 0.77/0.90 & 0.77/\textbf{0.92} \\
                                                & \textbf{VIE20} & 0.74/\textbf{0.87} & 0.68/0.86\\ \hline
        \multirow{3}{*}{\textbf{BUE}}           & \textbf{BLN}   & 0.62/\textbf{0.77} & 0.60/0.75 \\ 
                                                & \textbf{VIE14} & 0.79/0.90 & 0.82/\textbf{0.93} \\
                                                & \textbf{VIE20} & 0.72/0.88 & 0.77/\textbf{0.92} \\ 
    \end{tabular}
    \label{tab:comparing_results}
\end{table}

\section{Conclusion}

In this work, we propose two explainability visualization techniques tailored for transformers in FCM data.
The attention visualization helps to understand which cell population the model focuses on.
For instance, the fact that the model can identify meaningful biological structures in the data without being explicitly instructed to do so, suggests that it has a deep understanding of the data modality and is not simply relying on learned shortcuts for prediction. 
The gradient-based visualization allows to identify learned relationships between changes in the prediction and changes in the input data. 
By utilizing this approach, we could identify an overfitting behavior of the model for predicting \textit{Singlet} gates in B-ALL samples.
The proposed techniques are useful tools for assessing and debugging attention-based deep learning models for FCM data.
Future work could use the proposed interpretability techniques to introduce an inductive bias by imposing gradient-based regularization term in model training similar to \cite{mahapatra2022interpretability} or \cite{simard1998transformation}.

\section{Acknowledgement}
\label{sec: acknowledgement}
We thank Dieter Printz (FACS Core Unit, CCRI) for flow-cytometer maintenance and quality control, as well as Daniela Scharner and Susanne Suhendra-Chen (CCRI), Jana Hofmann (Charité), Mariann eDunken (HELIOS Klinikum), Marianela Sanz, Andrea Bernasconi, and Raquel Mitchell (Hospital Garrahan) for excellent technical assistance. We are indebted to Melanie Gau, Roxane Licandro, Florian Kleber, Paolo Rota and Guohui Qiao (all from TU Vienna) for valuable contributions to the AutoFLOW project. We thank Markus Kaymer and Michael Kapinsky (both from Beckman Coulter Inc.) for kindly assisting in the provision of customized DuraCloneTm tubes for this study as designed by the authors. Notably, Beckman Coulter Inc. did not have any influence on study design, data acquisition and interpretation, or manuscript writing. The study has received funding from the European Union’s H2020 Research and Innovation Program through Grant number 825749 “CLOSER: Childhood Leukemia: Overcoming Distance between South America and Europe Regions”, the Vienna Business Agency under grant agreement No 2841342 (Project MyeFlow) and by the Marie Curie Industry Academia Partnership \& Pathways (FP7-MarieCurie-PEOPLE-2013-IAPP) under grant no. 610872 to project “AutoFLOW” to MND. The authors acknowledge TU Wien Bibliothek for financial support through its Open Access Funding Programme.

\section{Declarations of interest}
Michael N. Dworzak received payments for travel, accommodation or other expenses from Beckman-Coulter. The other authors declare no competing financial interests.

\newpage
\bibliographystyle{unsrt}  
\bibliography{references}

\section{Appendix}

\subsection{Explicit Smoothing}
In \cite{smilkov2017smoothgrad} Smilkov et al. proposed \textbf{SmoothGrad}, an extension of the standard Saliency Map that aims to reduce visual noise in Saliency Maps by averaging the gradients of multiple randomly noised versions of the same input image and thereby smooths out local permutations.
We observed that the same procedure leads to more paralleled gradients in FCM samples. We, therefore, add noise $s \sim \mathcal{U}(0, 0.1)$ to the events and compute gradients. This action is repeated 10 times (empirically determined) and the average of the gradients is plotted.

\subsection{Qualitatively Evaluation of Attention heads}

In Equation \ref{eg:quant_blast_head} we state how the attendance of individual heads is measured.Cancer vs. non-cancer classification is a high imbalanced classification problem and we mind for this characteristic by measuring both the the proportion of cancer cells within the top 5\% to all cells in the top 5\% as well as the proportion of cancer cells within in the top5\% to all cancer cells.
This metric evaluates to $0$ if no cancer cells are among the top 5\% and it returns $1$ if either all cancer cells are among the top 5\% or if the top5\% solely consists of cancer.

We applied this metric only to calculate amount of cancer cells not any other biological structure in the FCM samples.
Expert generated ground truth is solely available for the binary classification cancer cells vs. non-cancer cells. Therefore, we could only qualitatively evaluate if our observation holds for other biological structures except cancer cells.

\subsection{Acute Myeloid Leukemia (AML)}
\begin{figure}
    \centering   
    \label{fig:aml_grad_step}
    \includegraphics[width=0.8\textwidth,keepaspectratio]{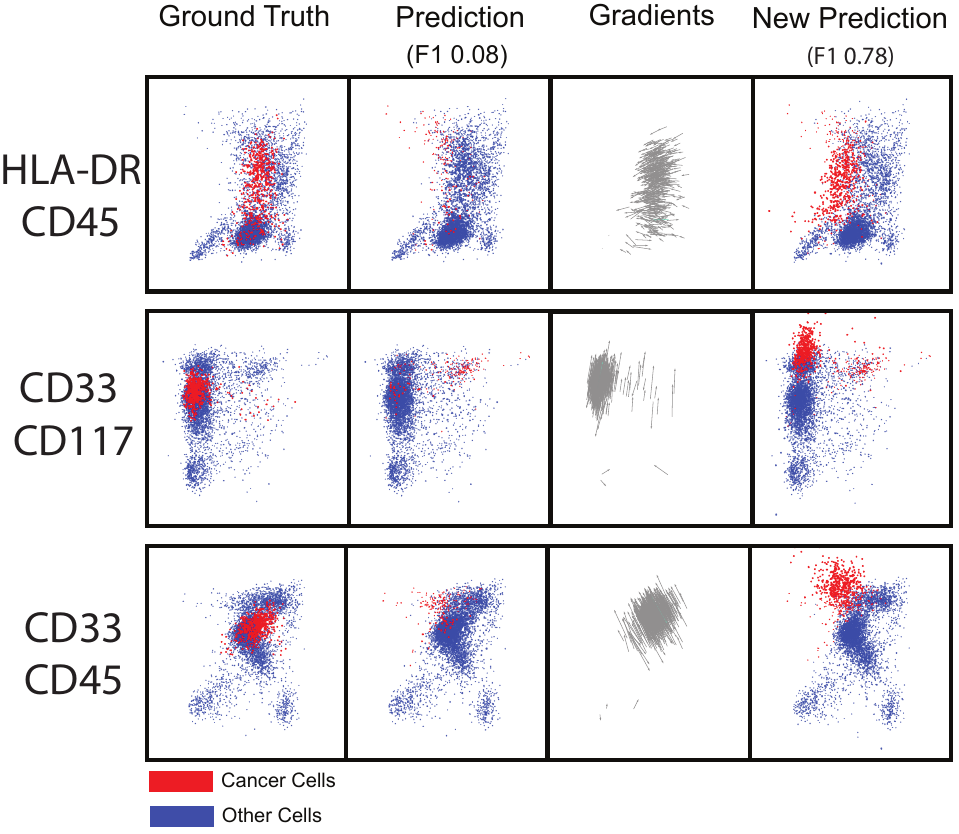}
     \caption{Here we demonstrate how the gradients of cell classification can be used to analyze a miss-classified AML FCM sample. When the false-negative cells are alternated in the direction of the gradients, a change in class predictions occurs. Each row depicts the same FCM sample from different 2D projections. The first two columns show the ground truth as well as the predict class-labels per event. The model hardly detects any cancer cells and reaches an F1-score below 0.1. In the third column. we see the top 500 biggest gradients of the summed classification output of the false-negative predicted cancer cells with respect to the input data. As shown in the 4th column, alternating these events in the direction of the gradients results in an F1-score improvement of over 60\%, as most of the false-negative predictions are now classified as cancer cells. This example demonstrates that the model does not anticipate cancer cells to be present at their previous location, since nearest change in class prediction is obtained by moving the cells to a position, which is more characteristic of common AML types. }
     
\end{figure}

In Figure \ref{fig:aml_grad_step} we showcase a practical application of using cell classification gradients to analyze a misclassified FCM sample of AML.
This example illustrates an interesting insight: our model doesn't anticipate cancer cells to be present at their original location. Instead, it achieves better class predictions by moving the cells to positions more characteristic of common AML types.

\end{document}